\documentclass[prd,preprintnumbers,nofootinbib]{revtex4} 
\usepackage{graphicx} 
\usepackage{amsmath}
\usepackage{amsfonts,amsbsy}
\usepackage{amssymb}

\def\be{\begin{equation}}
\def\ee{\end{equation}}
\def\bea{\begin{eqnarray}}
\def\eea{\end{eqnarray}}
\def\Tr{{\rm tr \;}}

\begin{document}

\title{\bf Forward dijets in high-energy collisions: evolution of QCD $n$-point 
functions beyond the dipole approximation}

\preprint{BCCUNY-HEP/10-02}
\preprint{RBRC-854}

\author{Adrian Dumitru$^{a,b,c}$ and Jamal Jalilian-Marian$^{b,c}$}
\affiliation{
$^a$ RIKEN BNL Research Center, Brookhaven National
  Laboratory, Upton, NY 11973, USA\\
$^b$ Department of Natural Sciences, Baruch College, CUNY,
17 Lexington Avenue, New York, NY 10010, USA\\
$^c$ The Graduate School and University Center, City
  University of New York, 365 Fifth Avenue, New York, NY 10016, USA}

\begin{abstract}
\noindent 
Present knowledge of QCD $n$-point functions of Wilson lines at high
energies is rather limited. In practical applications, it is therefore
customary to factorize higher $n$-point functions into products of
two-point functions (dipoles) which satisfy the BK evolution
equation. We employ the JIMWLK formalism to derive explicit evolution
equations for the 4- and 6-point functions of fundamental Wilson
lines and show that if the Gaussian approximation is carried out
before the rapidity evolution step is taken, then many leading order
$N_c$ contributions are missed. Our evolution equations could
specifically be used to improve calculations of forward dijet angular
correlations, recently measured by the STAR collaboration in
deuteron-gold collisions at the RHIC collider. Forward dijets in
proton-proton collisions at the LHC probe QCD evolution at even
smaller light-cone momentum fractions. Such correlations may provide
insight into genuine differences between the JIMWLK and BK approaches.
\end{abstract}

\maketitle

\section{Introduction}

The Jalilian-Marian$-$Iancu$-$McLerran$-$Weigert$-$Leonidov$-$Kovner
(JIMWLK) functional evolution equation describes the energy dependence
of $n$-point functions of Wilson lines at small light-cone momentum
fraction $x$~\cite{Balitsky:1995ub,jimwlk}. These $n$-point functions
appear in multi-particle production cross sections in hadronic (or
heavy-ion) collisions. To date, our knowledge of the behavior of such
$n$-point functions in QCD is very limited. Therefore, it is common to
employ a Gaussian (and large-$N_c$) approximation which reduces these
functions to powers of the two-point function, which at high
transverse momentum corresponds to the well-known
Balitsky-Fadin-Kuraev-Lipatov (BFKL) unintegrated gluon
distribution~\cite{bfkl}. The evolution of the two-point function in
the dipole approximation~\cite{dipole_model} is determined by an
ordinary integro-differential equation known as the Balitsky-Kovchegov
(BK) equation~\cite{Balitsky:1995ub,Kovchegov:1999yj} which has been
the subject of intense theoretical investigation in the past few
years.

\begin{figure}[hb]
\begin{center}
\includegraphics[width=7.5cm]{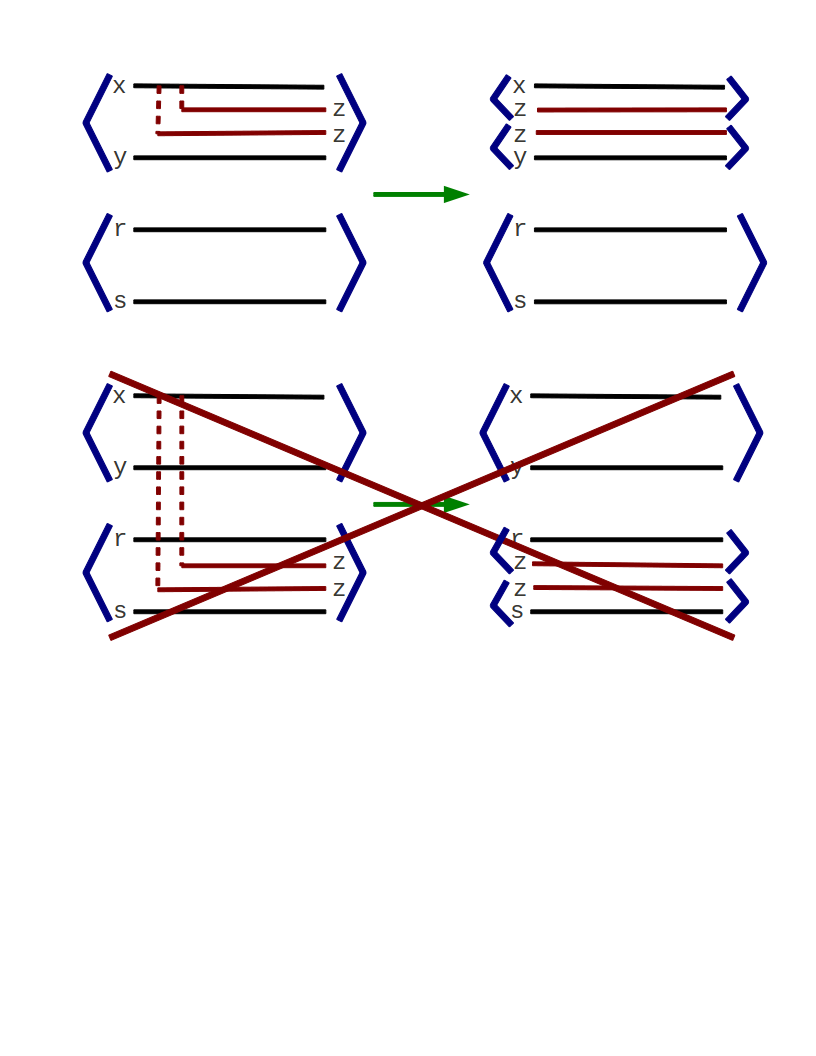}
\hspace{20mm} 
\includegraphics[width=7.5cm]{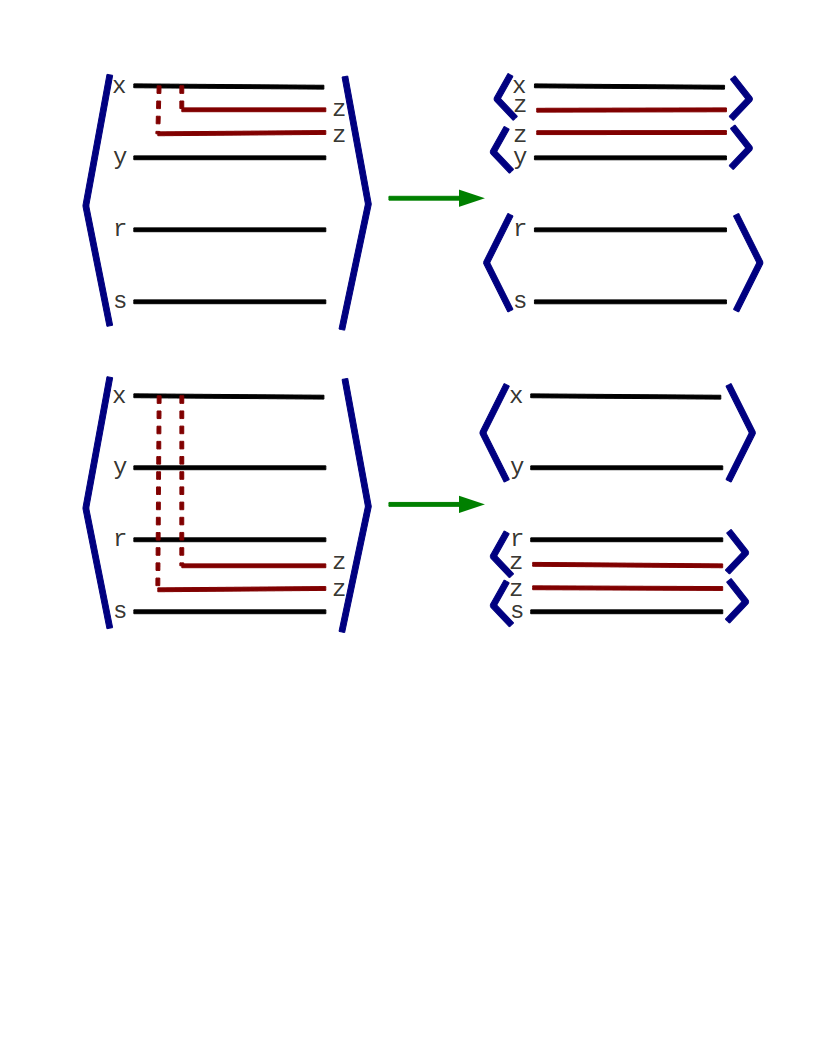}
\end{center}
\vspace*{-4cm}
\caption[a]{Left: BK-evolution of a set of dipoles: in an evolution
  step, emission of a gluon can only split the parent dipole. Right:
  JIMWLK evolution of the $n$-point function allows for
  additional contributions.}
\label{fig:RadGaussJIMWLK}
\end{figure}

Evolution with energy or rapidity $y\sim\log 1/x$ occurs by (real or
virtual) radiation of an additional gluon from a given $n$-point
operator. Cross sections are related to expectation values of traces
of such operators which project onto their physical matrix
elements. If the expectation value of the $n$-point operator is split
into dipoles before the radiation of the additional gluon (in order to
perform the evolution step by means of the BK equation) then only the
dipole from which the radiation emerged is allowed to split into two
dipoles. On the other hand, if the evolution step is given by the
JIMWLK equation then additional contributions arise, even at leading
order in $N_c$. This is illustrated in fig.~\ref{fig:RadGaussJIMWLK}. 

To exhibit the differences between Gaussian+BK and JIMWLK evolution
more clearly we focus on two specific $n$-point functions which appear
in (forward) quark+gluon dijet production in pA
collisions. Hadron-nucleus collisions are well suited for
investigations of high gluon density QCD due to the fact that
one can derive analytic relations for particle production when only
the target is dense, while the projectile is dilute. Furthermore,
hadron-nucleus collisions are free of the hot final state medium
(``quark-gluon plasma'') produced in AA collisions which may affect
the observables. Very recently, the STAR collaboration has presented
two-hadron angular correlations in the forward rapidity region of
deuteron-gold collisions which show a weakening/disappearance of the
away side peak~\cite{Braidot:2010zh}, in agreement with expectations
from gluon saturation
dynamics~\cite{Kharzeev:2004bw,Frankfurt:2007rn,Tuchin:2009nf,Albacete:2010pg}.
The purpose of the present paper is to derive explicit and complete
(in terms of $N_c$ counting) evolution equations for the relevant
$n$-point functions so that quantitative theoretical expectations
could eventually be obtained.

The cross section for production of a valence quark plus a gluon at
forward rapidity in pA collisions was calculated in \cite{jjmyk1} (in
momentum space) and in \cite{Marquet:2007vb} (in coordinate space); we
also refer the reader to refs.~\cite{Other_qg}. For completeness, we
reproduce the expressions in the appendix. They involve expectation
values of products of (up to) six Wilson lines in the fundamental
representation (single-inclusive production involves only the
$2-$point function \cite{adjjm}).  Explicitly, the following operators
appear in the two parton (quark+gluon) production cross section
\be
O_4 (r, \bar{r}:s) \equiv \Tr V^\dagger_r\, t^a \, V_{\bar{r}}\, t^b
\, [U_s]^{ab}  = 
{1\over 2} \bigg[ \Tr V_r^\dagger\, V_s ~\, \Tr V_{\bar{r}} \, V_s^\dagger - 
{1\over N_c} \Tr V_r^\dagger \, V_{\bar{r}} \bigg]
\label{eq:o_4}
\ee
and 
\be
O_6 (r,\bar{r}:s,\bar{s}) \equiv \Tr V_r \, V^\dagger_{\bar{r}}\,
t^a\, t^b \, [U_s\, U^\dagger_{\bar{s}}]^{ba} =
{1\over 2} \bigg[
\Tr V_r \, V^\dagger_{\bar{r}}\, V_{\bar{s}} \, V^\dagger_s
~ \, \Tr V_s \, V^\dagger_{\bar{s}} 
- {1\over N_c} \Tr V_r \, V^\dagger_{\bar{r}}\bigg]
\label{eq:o_6}
\ee
where $V$ ($U$) is a Wilson line in the fundamental (adjoint)
representation and $r$, $s$ etc.\ denote two-dimensional
coordinates in the transverse plane. Here, we have used the identity
\be
U^{ab}\, t^b = V^\dagger \, t^a\, V
\label{eq:a2f}
\ee
to relate matrices in the two representations.

Due to the fact that explicit evolution equations for the above
combinations of Wilson lines have not been derived so far, and therefore
solutions to these equations are unknown, it is common to resort to
the large $N_c$ and Gaussian
approximation~\cite{Albacete:2010pg,Marquet:2007vb,Blaizot:2004wv,Fukushima:2007dy}. In
this approximation, these two expectation values can be written as
\bea \langle O_4 (r, \bar{r}:s)  \rangle \; &\simeq& \; \langle O_2 (r - s)\rangle \; \langle
O_2 (s - \bar{r})\rangle  \nonumber \\ 
\langle O_6 (r,\bar{r}:s,\bar{s}) 
\rangle \; &\simeq& \; 
\langle O_2 (r - s) \rangle \; \langle O_2 (\bar{r} - \bar{s}) \rangle \; \langle O_2
(s - \bar{s}) \rangle  + 
\langle O_2 (r - \bar{r}) \rangle \; \langle O_2 (\bar{s} - s) \rangle \; \langle O_2
(s - \bar{s}) \rangle 
\label{eq:o_gauss}
\eea 
where $\langle O_2 (r, \bar{r})\rangle \equiv \langle \Tr V_r \,
V^\dagger_{\bar{r}} \rangle$. To keep track of factors of $N_c$ it is
convenient to normalize expectation values of traces of Wilson lines,
i.e.\ the S matrices, as follows:
\bea
S (r, \bar{r}) &\equiv& {1\over C_A} \, \langle O_2 \rangle \nonumber \\
S_4 (r, \bar{r}: s)&\equiv& {1\over C_A\, C_F} \, \langle O_4 \rangle
\nonumber \\
S_6 (r, \bar{r}: s, \bar{s}) &\equiv& {1\over C_A\, C_F} \, \langle
O_6 \rangle~.
\label{eq:norm_s}
\eea
For phenomenological applications of the Color Glass Condensate
formalism to two-particle production, it is common to employ the
approximations~(\ref{eq:o_gauss}) to factorize the higher point
functions of Wilson lines into products of two point functions.  This
greatly simplifies applications as it allows one to write $S_4$ and
$S_6$ in terms of the well known BK two point function. Then BK evolution
of the two point function $S$ is assumed to account for small-$x$
evolution of the higher point functions $S_4$ and $S_6$
\cite{Albacete:2010pg}. We shall show below that indeed this procedure
retains the leading-$N_c$ contribution to the $4$-point function
$S_4$. On the other hand, we also show that it misses many
leading-$N_c$ contributions to the evolution of the $6$-point function
$S_6$.

\section{Evolution equations for higher point functions}

In this section we derive explicit evolution equations for the
expectation values of $O_4$ and $O_6$ which appear in the two-particle
production cross section. We then apply the large-$N_c$ and Gaussian
approximations to our evolution equations and compare to the ``naive''
result~(\ref{eq:o_gauss}) where BK evolution for the two point
functions is used on the right hand side of~(\ref{eq:o_gauss}).

We start from the JIMWLK evolution equation which determines the small-$x$
evolution of any $n$-point function $O$ from
\be
\frac{d}{dy} \langle O \rangle =
  \frac{1}{2} \left< \int d^2x \, d^2y \, \frac{\delta}{\delta\alpha_x^b}
     \, \eta^{bd}_{xy} \, \frac{\delta}{\delta\alpha_y^d} \, O \right>~,
   \label{eq:ham}
\ee
where
\be  \label{eq:eta}
\eta^{bd}_{xy} = \frac{1}{\pi} \int \frac{d^2z}{(2\pi)^2}
     \frac{(x-z)\cdot(y-z)}{(x-z)^2 (y-z)^2} \left[
       1 + U^\dagger_x U_y - U^\dagger_x U_z - U^\dagger_z U_y
       \right]^{bd} ~.
\ee
We first focus on $O_4$ defined in eq.~(\ref{eq:o_4}). The second term 
(the two point function) from that equation evolves according to the JIMWLK
equation 
\be
{d\over dy} \langle \Tr V_r^\dagger \, V_{\bar{r}} \rangle = -
{N_c\, \alpha_s \over 2\pi^2} \int\,
d^2 z\, {(r - \bar{r})^2 \over (r - z)^2 (\bar{r} - z)^2} \, \left<
\Tr V_r^\dagger \, V_{\bar{r}} - {1\over N_c} \Tr V_r^\dagger \, V_z~
\Tr\, V_{\bar{r}} \, V^\dagger_z\right>~.
\label{eq:2pt}
\ee 
This reduces to the BK equation in the Gaussian and large-$N_c$
approximations. On the other hand, the first term in $O_4$ involving
four fundamental Wilson lines satisfies
\be
{d\over dy} \langle \Tr V_r^\dagger\, V_s~\, \Tr V_{\bar{r}} \,
V_s^\dagger  \rangle = 
\frac{1}{2} \left< \int d^2x \, d^2y \, \frac{\delta}{\delta\alpha_x^b}
     \, \eta^{bd}_{xy} \, \frac{\delta}{\delta\alpha_y^d}  \,
\Tr V_r^\dagger\, V_s~  \Tr V_{\bar{r}} \, V_s^\dagger \right> ~.
\label{eq:4pt}
\ee
Using the explicit form of $\eta^{bd}_{xy}$ given by (\ref{eq:eta}),
we note that the derivative of $\eta$ (w.r.t.\ $\alpha_x^b$) vanishes
due to the color structure, except for the last term $\sim U^\dagger_z
U_y $ where one does need to differentiate $U_y$.  The calculation is
straightforward but lengthy and involves repeated use of
eq.~(\ref{eq:a2f}) as well as the Fierz identity for the product of
fundamental matrices
\be
[t^a]_{ij}\, [t^a]_{kl} = {1\over 2} \bigg[\delta_{il}\, \delta_{jk} -
  {1\over N_c} \delta_{ij} \delta_{kl}\bigg]~.
\label{eq:fierz}
\ee
Adding the evolution equation for the two point function, eq.~(\ref{eq:2pt}),
we obtain
\bea
&&{d\over dy} \langle O_4 (r, \bar{r}: s) \rangle  =  
- {N_c\, \alpha_s \over (2\pi)^2} 
\int d^2 z \, \Biggl< 
\bigg[{(r - s)^2 \over (r - z)^2 (s - z)^2} + {(\bar{r} - s)^2 \over
    (\bar{r} - z)^2 (s - z)^2 }\bigg] \,
\Tr V_r^\dagger \, V_s ~ \Tr V^\dagger_s\, V_{\bar{r}} \nonumber \\
&+& {1\over N_c} \Bigg[
 {1\over 2} \bigg[ {(r - s)^2 \over (r - z)^2 (s - z)^2} + {(\bar{r} -
     s)^2 \over (\bar{r} - z)^2 (s - z)^2 }
- {(r - \bar{r})^2 \over (r - z)^2 (\bar{r} - z)^2 } \bigg] 
\bigg[\Tr V_r^\dagger \, V_z \, V^\dagger_s\, V_{\bar{r}} \, V_z^\dagger \, V_s +  
\Tr V_r^\dagger \, V_s \, V^\dagger_z\, V_{\bar{r}} \, V_s^\dagger \,
V_z\bigg] \nonumber \\ & -&
 {(r - s)^2 \over (r - z)^2 (s - z)^2}\, \Tr V_r^\dagger \, V_z ~\,
\Tr  V^\dagger_s\, V_{\bar{r}} ~\, \Tr V_z^\dagger \, V_s 
- {(\bar{r} - s)^2 \over (\bar{r} - z)^2 (s - z)^2 }
\Tr V_r^\dagger \, V_s ~\, \Tr V^\dagger_z\, V_{\bar{r}} ~\, 
\Tr\, V_s^\dagger \, V_z \nonumber \\
&-& \bigg[{(r - s)^2 \over (r - z)^2 (s - z)^2} + {(\bar{r} - s)^2
    \over (\bar{r} - z)^2 (s - z)^2 }\bigg] \,
\Tr V_r^\dagger \, V_{\bar{r}}\, \Bigg] \nonumber \\
&+&  {1\over N_c^2} \, {(r - \bar{r})^2 \over (r - z)^2 (\bar{r} -
  z)^2} \, \Tr\, V_r^\dagger \, V_z  ~\, \Tr\, V_z^\dagger \, V_{\bar{r}}
\;\Biggr >~.
\label{eq:ev_o_4}
\eea
Note that the first and fourth line on the rhs could be combined into
$O_4$ by virtue of the identity~(\ref{eq:o_4}); thus, all terms on the
rhs feature at least as many Wilson lines as the original operator.

Equation~(\ref{eq:ev_o_4}) describes the small-$x$ (rapidity) evolution of the
$4$-point function $O_4$. It has the intuitive interpretation of the
scattering matrix for two quark-anti-quark dipoles evolving with
rapidity while (multiply) scattering from the target. It is worth
noting that this equation remains finite when the internal integration variable 
$z$ approaches any of the external coordinates $r$, $\bar{r}$, $s$.

We now apply the large $N_c$ and Gaussian approximations to
eq.~(\ref{eq:ev_o_4}). The leading-$N_c$ contributions on the rhs
originate from the first and third lines of~(\ref{eq:ev_o_4}). Written
in terms of the scattering matrix $S_4$,
\bea
{d\over dy} S_4  &\simeq &
- {N_c\, \alpha_s \over 2\pi^2} \;\; 
\int d^2 z \; \Bigg\{
S (s - \bar{r}) \, {(r - s)^2 \over (r - z)^2 (s - z)^2}\,
\bigg[S (r -s) -  S (r -z) \, S (z - s) \bigg]
\nonumber \\
& & \hspace{2.4cm} +\;
S (r - s ) \, {(\bar{r} - s)^2 \over (\bar{r} - z)^2 (s - z)^2}\,
\bigg[S (\bar{r} -s) -  S (\bar{r} -z) 
\, S (z - s) \bigg] \Bigg\} ~.
\label{eq:S_4_gauss}
\eea
Using the BK equation for the $2$-point function $S$,
\be
{d\over dy} S (r - s)  = 
- {N_c\, \alpha_s \over 2\pi^2} \;
\int d^2 z \,
 {(r - s)^2 \over (r - z)^2 (s - z)^2}\,\bigg[S (r -s) -
  S (r -z) \, S (z - s) \bigg]
\label{eq:bk}
\ee
we can rewrite eq.~(\ref{eq:S_4_gauss}) as
\be
{d\over dy} S_4 (r, \bar{r}: s) \simeq  S (s - \bar{r}) \, {d\over
  dy} S (r - s) +
S (r - s ) \, {d\over dy} S (s - \bar{r})~.
\label{eq:S_4_gauss_simp}
\ee
Thus, it is evident that the factorized form~(\ref{eq:o_gauss}) for
the $4$-point function $\langle O_4\rangle$ correctly captures the
leading-$N_c$ contributions. In passing, we note that to arrive at
eq.~(\ref{eq:S_4_gauss_simp}) we have dropped 12 terms of order
$1/N_c^2$.

Next, we derive the evolution equation for the $6$-point function
$O_6$ from the JIMWLK equation~(\ref{eq:ham}). The procedure is 
straightforward but very tedious, here we just quote the final
result:
\bea
{d\over dy}\!\!\!
&& \!\!\!\!\! \langle O_6 (r,\bar{r}:s,\bar{s}) \rangle
 =  - {N_c\, \alpha_s \over 2(2\pi)^2} \int d^2 z \nonumber \\
&&
\Bigg< \left[
{(r - s)^2 \over (r - z)^2 (s - z)^2} +  {(r - \bar{r})^2 \over (r -
  z)^2 (\bar{r} - z)^2}
+  {(\bar{r} - \bar{s})^2 \over (\bar{r} - z)^2 (\bar{s} - z)^2}
+   3 {(s - \bar{s})^2 \over (s - z)^2 (\bar{s} - z)^2}\right]\, 
\Tr V_r \, V^\dagger_{\bar{r}}\, V_{\bar{s}} \, V^\dagger_s~\, 
\Tr V_s \, V^\dagger_{\bar{s}} \nonumber \\
&+& {1\over N_c} 
\Bigg[
- \big[ - {(r - s)^2 \over (r - z)^2 (s - z)^2} + {(r - \bar{s})^2
    \over (r - z)^2 (\bar{s} - z)^2} -
{(s - \bar{s})^2 \over (s - z)^2 (\bar{s} - z)^2}\big] \,
\Tr V_r \, V^\dagger_{\bar{r}}\, V_{\bar{s}} \, V^\dagger_s\,  V_z \,
V^\dagger_{\bar{s}}\, V_s \, V^\dagger_z 
\nonumber \\
&& -  \big[ {(\bar{r} - s)^2 \over (\bar{r} - z)^2 (s - z)^2} -
  {(\bar{r} - \bar{s})^2 \over (\bar{r} - z)^2 (\bar{s} - z)^2} - {(s
    - \bar{s})^2 \over (s - z)^2 (\bar{s} - z)^2}\big]\,
\Tr V_r \, V^\dagger_{\bar{r}}\, V_z \, V^\dagger_{\bar{s}} \, V_s\,
V^\dagger_z \, V_{\bar{s}}\, V^\dagger_s
\nonumber\\
&& -  \big[{(\bar{r} - s)^2 \over (\bar{r} - z)^2 (s - z)^2} -
  {(\bar{r} - \bar{s})^2 \over (\bar{r} - z)^2 (\bar{s} - z)^2} - {(r
    - s)^2 \over (r - z)^2 (s - z)^2} +  {(r - \bar{s})^2 \over (r -
    z)^2 (\bar{s} - z)^2} \big]\,
\Tr V_r \, V^\dagger_z\, V_s \, V^\dagger_{\bar{s}} \, V_z\,
V^\dagger_{\bar{r}} \, V_{\bar{s}}\, V^\dagger_s
\nonumber \\
&&
+ 2 {(s - \bar{s})^2 \over (s - z)^2 (\bar{s} - z)^2} \,
\Tr V_r \,V^\dagger_{\bar{r}}\, V_{\bar{s}} \, V^\dagger_z \, V_s\,
V^\dagger_{\bar{s}} \, V_z\, V^\dagger_s
\nonumber \\
&& 
+ \big[{(\bar{r} - s)^2 \over (\bar{r} - z)^2 (s - z)^2} - {(\bar{r} -
    \bar{s})^2 \over (\bar{r} - z)^2 (\bar{s} - z)^2} - {(r - s)^2
    \over (r - z)^2 (s - z)^2} +  {(r - \bar{s})^2 \over (r - z)^2
    (\bar{s} - z)^2} \big]\,
\Tr V_r \, V^\dagger_{\bar{s}} \, V_s\,  V^\dagger_{\bar{r}} \,
V_{\bar{s}}\, V^\dagger_s
\nonumber\\
&&
-\big[ {(r - \bar{r})^2 \over (r - z)^2 (\bar{r} - z)^2} + {(r - s)^2
    \over (r - z)^2 (s - z)^2} -
{(s - \bar{r})^2 \over (s - z)^2 (\bar{r} - z)^2}\big] \,
\Tr V_z \, V^\dagger_{\bar{r}} \, V_{\bar{s}} \, V^\dagger_s ~\, 
\Tr V_r \, V^\dagger_z ~\,
\Tr V_s \, V^\dagger_{\bar{s}} \nonumber \\
&&
-  \big[ {(r - \bar{r})^2 \over (r - z)^2 (\bar{r} - z)^2} + {(\bar{r}
    - \bar{s})^2 \over (\bar{r} - z)^2 (\bar{s} - z)^2} - {(r -
    \bar{s})^2 \over (r - z)^2 (\bar{s} - z)^2}\big]\,
\Tr V_r \, V^\dagger_z \, V_{\bar{s}} \, V^\dagger_s ~\, 
\Tr V_z \, V^\dagger_{\bar{r}} ~\, 
\Tr V_s \, V^\dagger_{\bar{s}} \nonumber \\
&&     
-  \big[ {(r - s)^2 \over (r - z)^2 (s - z)^2} + {(s - \bar{s})^2
    \over (s - z)^2 (\bar{s} - z)^2} - {(r - \bar{s})^2 \over (r -
    z)^2 (\bar{s} - z)^2}\big]\,
\Tr V_r \, V^\dagger_{\bar{r}} \, V_{\bar{s}} \, V^\dagger_z ~\, 
\Tr V_z \, V^\dagger_s ~\, 
\Tr V_s \, V^\dagger_{\bar{s}} \nonumber \\
&&     
-  \big[ {(\bar{r} - \bar{s})^2 \over (\bar{r} - z)^2 (\bar{s} - z)^2}
  + {(s - \bar{s})^2 \over (s - z)^2 (\bar{s} - z)^2} - {(\bar{r} -
    s)^2 \over (\bar{r} - z)^2 (s - z)^2}\big]\,
\Tr V_r \, V^\dagger_{\bar{r}} \, V_z \, V^\dagger_s ~\, 
\Tr V_{\bar{s}} \, V^\dagger_z ~\, 
\Tr V_s \, V^\dagger_{\bar{s}} \nonumber \\
&&     
-2  {(s - \bar{s})^2 \over (s - z)^2 (\bar{s} - z)^2} \,
\Tr V_r \, V^\dagger_{\bar{r}} \, V_{\bar{s}} \, V^\dagger_s ~\, 
\Tr V_s \, V^\dagger_z \, 
\Tr V_z \, V^\dagger_{\bar{s}} \nonumber \\
&&    
- \big[{(\bar{r} - s)^2 \over (\bar{r} - z)^2 (s - z)^2} - {(\bar{r}
- \bar{s})^2 \over (\bar{r} - z)^2 (\bar{s} - z)^2}
- {(r - s)^2 \over (r - z)^2 (s - z)^2} +  {(r - \bar{s})^2 \over (r
 - z)^2 (\bar{s} - z)^2} \big]\,
\Tr V_r \, V^\dagger_s ~\, 
\Tr\, V^\dagger_{\bar{r}}\, V_{\bar{s}} ~\, 
\Tr V_s \, V^\dagger_{\bar{s}} \nonumber \\
&&     
+  \big[- {(\bar{r} - s)^2 \over (\bar{r} - z)^2 (s - z)^2} - {(r -
    \bar{s})^2 \over (r - z)^2 (\bar{s} - z)^2}
+ {(r - \bar{r})^2 \over (r - z)^2 (\bar{r} - z)^2} +  {(s -
  \bar{s})^2 \over (s - z)^2 (\bar{s} - z)^2} \big]\,
\Tr V_r \, V^\dagger_{\bar{r}} ~\, 
\Tr V_{\bar{s}}\, V^\dagger_s ~\, 
\Tr V_s \, V^\dagger_{\bar{s}} \nonumber \\
&&   
+ \big[{(\bar{r} - s)^2 \over (\bar{r} - z)^2 (s - z)^2} + {(r -
    \bar{s})^2 \over (r - z)^2 (\bar{s} - z)^2}
- {(\bar{r} - \bar{s})^2 \over (\bar{r} - z)^2 (\bar{s} - z)^2} -  {(r
  - s)^2 \over (r - z)^2 (s - z)^2}
-2 {(r - \bar{r})^2 \over (r - z)^2 (\bar{r} - z)^2} \nonumber \\
&&
-4 {(s - \bar{s})^2 \over (s - z)^2 (\bar{s} - z)^2} \big]\,
\Tr V_r \, V^\dagger_{\bar{r}}\Bigg] \nonumber \\
&+&
{2\over N_c^2}  {(r - \bar{r})^2 \over (r - z)^2 (\bar{r} - z)^2} \,
\Tr V_r\, V_z ~\, 
\Tr V^\dagger_{\bar{r}} \, V^\dagger_z  
\Bigg>
\label{eq:ev_o_6}
\eea
We have checked that this equation remains finite when the internal
integration variable $z$ approaches any of the external coordinates
$r$, $\bar{r}$, $s$, $\bar{s}$. Note the appearance of lower point
functions on the right hand side of the equation. Unlike the similar
terms in the evolution of $O_4$ these can not be combined into the
 original operator $O_6$.

We may now employ the Gaussian approximation on the rhs of the
equation above to exhibit the leading-$N_c$ contributions (from lines
$1$ and $7-13$). The equation reduces to
\bea
&&{d\over dy}  S_6
\simeq    - {N_c\, \alpha_s \over (2\pi)^2} \int \, d^2 z \, \Bigg\{ 
\bigg[{(r - s)^2 \over (r - z)^2 (s - z)^2} +  {(r - \bar{r})^2 \over
    (r - z)^2 (\bar{r} - z)^2}
+  {(\bar{r} - \bar{s})^2 \over (\bar{r} - z)^2 (\bar{s} - z)^2}
+   3 {(s - \bar{s})^2 \over (s - z)^2 (\bar{s} - z)^2}\bigg]\nonumber \\ 
&& \hspace{4cm} \bigg[S (r - \bar{r})\, S (s - \bar{s}) \, S (s - \bar{s}) + 
 S (r - s)\, S (\bar{r} - \bar{s}) \, S (s - \bar{s}) \bigg] \nonumber \\
&+&\bigg[ -{(r - \bar{r})^2 \over (r - z)^2 (\bar{r} - z)^2} - {(r -
    s)^2 \over (r - z)^2 (s - z)^2} +
{(s - \bar{r})^2 \over (s - z)^2 (\bar{r} - z)^2}\bigg] 
\bigg[\bigg( S (z - \bar{r})\, S (s - \bar{s}) \, +  S (z - s)\, S
  (\bar{r} - \bar{s})
\bigg)\, S (r - z) \, S (s - \bar{s})\bigg]  \nonumber \\
&+& \bigg[{(r - \bar{s})^2 \over (r - z)^2 (\bar{s} - z)^2} - {(r -
    \bar{r})^2 \over (r - z)^2 (\bar{r} - z)^2} - {(\bar{r} -
    \bar{s})^2 \over (\bar{r} - z)^2 (\bar{s} - z)^2}\bigg]
\bigg[\bigg( S (r - z)\, S (s - \bar{s}) \, +  S (r - s)\, S (z - \bar{s}) 
\bigg)\, S (z - \bar{r}) \, S (s - \bar{s})\bigg]  \nonumber \\
&-&  \bigg[ {(r - s)^2 \over (r - z)^2 (s - z)^2} + {(s - \bar{s})^2
    \over (s - z)^2 (\bar{s} - z)^2} - {(r - \bar{s})^2 \over (r -
    z)^2 (\bar{s} - z)^2}\bigg]
\bigg[\bigg( S (r - \bar{r})\, S (\bar{s} - z) + S (r - z)\, S
  (\bar{r} - \bar{s}) \bigg) 
 S (z - s) \, S (s - \bar{s})\bigg] \nonumber \\
&-&
\bigg[ {(\bar{r} - \bar{s})^2 \over (\bar{r} - z)^2 (\bar{s} - z)^2} +
  {(s - \bar{s})^2 \over (s - z)^2 (\bar{s} - z)^2} - {(\bar{r} - s)^2
    \over (\bar{r} - z)^2 (s - z)^2}\bigg]
\bigg[\bigg( S (r - \bar{r})\, S (z - s) + S (r - s)\, S (\bar{r} - z) \bigg)
 S (\bar{s} - z) \, S (s - \bar{s})\bigg] \nonumber \\
&-&2  {(s - \bar{s})^2 \over (s - z)^2 (\bar{s} - z)^2}
\bigg[\bigg( S (r - \bar{r})\, S (s - \bar{s}) + S (r - s)\, S
  (\bar{r} - \bar{s}) \bigg)
 S (s - z) \, S (z - \bar{s})\bigg] \nonumber \\
&-&
 \bigg[{(\bar{r} - s)^2 \over (\bar{r} - z)^2 (s - z)^2} - {(\bar{r} -
     \bar{s})^2 \over (\bar{r} - z)^2 (\bar{s} - z)^2}
- {(r - s)^2 \over (r - z)^2 (s - z)^2} +  {(r - \bar{s})^2 \over (r -
  z)^2 (\bar{s} - z)^2} \bigg]
S (r - s)\, S (\bar{r} - \bar{s})\, S (s - \bar{s})
\nonumber \\
&+&
\bigg[- {(\bar{r} - s)^2 \over (\bar{r} - z)^2 (s - z)^2} - {(r -
    \bar{s})^2 \over (r - z)^2 (\bar{s} - z)^2}
+ {(r - \bar{r})^2 \over (r - z)^2 (\bar{r} - z)^2} +  {(s -
  \bar{s})^2 \over (s - z)^2 (\bar{s} - z)^2} \bigg] S (r - \bar{r})\,
S (s - \bar{s})\, S (s - \bar{s})
\Bigg\}
\label{eq:s_6_gauss}
\eea

There are many more terms in this equation than obtained by 
differentiating eq.~(\ref{eq:o_gauss}). To make this more
clear, we write eq.~(\ref{eq:s_6_gauss}) in the form
\bea &&{d\over dy} S_6 (r, \bar{r}: s, \bar{s}) \simeq {d\over
dy} \bigg[S (r -s)\; S (\bar{s} - \bar{r})\; S (s - \bar{s}) \; +
    \; S (r -\bar{r})\; S (s - \bar{s})\; S (s - \bar{s})\bigg]
  \nonumber \\ &-& {N_c\, \alpha_s \over (2\pi)^2} \, S (s - \bar{s})
  \int \, d^2 z \, \Bigg\{\nonumber \\ && ~~ \bigg[{(r - s)^2 \over (r
      - z)^2 (s - z)^2} + {(\bar{r} - \bar{s})^2 \over (\bar{r} - z)^2
      (\bar{s} - z)^2} - {(\bar{r} - s)^2 \over (\bar{r} - z)^2 (s -
      z)^2} - {(r - \bar{s})^2 \over (r - z)^2 (\bar{s} - z)^2} \bigg]
  \nonumber \\ && \hspace{3cm} \bigg[S (r - \bar{r}) - S (z -
    \bar{r})\, S (r - z) \bigg] \, S (s - \bar{s}) \nonumber \\ && +
  \bigg[ {(r - \bar{r})^2 \over (r - z)^2 (\bar{r} - z)^2} + {(s -
      \bar{s})^2 \over (s - z)^2 (\bar{s} - z)^2} - {(\bar{r} - s)^2
      \over (\bar{r} - z)^2 (s - z)^2} - {(r - \bar{s})^2 \over (r -
      z)^2 (\bar{s} - z)^2} \bigg] \nonumber \\ && \hspace{3cm} \bigg[
    S (r - s) - S (z - s) \, S (r - z) \bigg] \, S (\bar{r} - \bar{s})
  \nonumber \\ && + \bigg[{(r - \bar{s})^2 \over (r - z)^2 (\bar{s} -
      z)^2} - {(r - \bar{r})^2 \over (r - z)^2 (\bar{r} - z)^2} - {(s
      - \bar{s})^2 \over (s - z)^2 (\bar{s} - z)^2} + {(\bar{r} - s)^2
      \over (\bar{r} - z)^2 (s - z)^2} \bigg] S (r - s)\, S (z -
  \bar{s}) \, S (z - \bar{r}) \nonumber \\ && - \bigg[ {(r - s)^2
      \over (r - z)^2 (s - z)^2} - {(r - \bar{s})^2 \over (r - z)^2
      (\bar{s} - z)^2} + {(\bar{r} - \bar{s})^2 \over (\bar{r} - z)^2
      (\bar{s} - z)^2} - {(\bar{r} - s)^2 \over (\bar{r} - z)^2 (s -
      z)^2} \bigg] S (r - \bar{r})\, S (\bar{s} - z) \, S (z - s)
  \Bigg\} \nonumber\\ & & 
\eea 
The terms in the bracket (first line) could also be obtained from the
Gaussian approximation~(\ref{eq:o_gauss}) to the $6$-point function
combined with BK evolution of the dipoles\footnote{The second term in
  the bracket in the first line seems to have been omitted in
  ref.~\cite{Marquet:2007vb}.}. All other terms would be missed. The
structure of the extra terms suggests why a naive Gaussian
approximation fails; to see this, consider $O_6$ from
eq.~(\ref{eq:o_6}). The first term is a product of two traces, a trace
of $4$ Wilson lines times a trace of two Wilson lines.  Emission of a
gluon between a different quark-anti-quark pair is missed by the
Gaussian approximation. In other words, if the Gaussian approximation
is applied to the trace of four Wilson lines in eq.~(\ref{eq:o_6})
before the rapidity evolution step, then due to color neutrality, a
radiated gluon can not end up in another dipole. On the other
hand, if one performs the rapidity evolution step of the full 4-point
function then the emitted gluon can end up anywhere between other
quark and anti-quark pairs, including those which are not the
``parents'' of the radiated gluon. These are precisely the
contributions which are missed by the Gaussian approximation in the
case of the $6$-point function\footnote{In this respect, the effects
  discussed here go beyond the correlations seen in dipole chains of
  the form $\Tr V_xV^\dagger_y\,~\Tr V_rV^\dagger_{\bar r}\,~\Tr
  V_sV^\dagger_{\bar s}\cdots$ \cite{DipChainCorr}. For such operators
  there is no cross-dipole emission as in
  fig.~\ref{fig:RadGaussJIMWLK}.}. It is easy to see that this is not
possible for the $4$-point function (\ref{eq:o_4}) since the traces
involve only two Wilson lines, and only one Gaussian contraction is
possible (at leading order in $N_c$). This also means that our
findings here do not affect fully inclusive observables such as DIS
structure functions $F_2$ and $F_L$, or single inclusive particle
production in pA collisions, since those involve the two point
function; its evolution is only sensitive to the $4$-point function
which does not receive any leading $N_c$ corrections from JIMWLK (as
compared to BK).

In summary, we have derived explicit evolution equations for the
$n$-point functions that appear in forward dijet angular correlations
in pA collisions. We find that factorizing these $n$-point functions
into dipoles {\em before} performing evolution in rapidity misses many
leading-$N_c$ contributions; higher multipole operators obey different
evolution equations which can not be reduced to BK evolution of
dipoles~\cite{multipole}. Moreover, a rather large number of
$N_c$-suppressed terms arises in the full JIMWLK evolution equations
for higher $n$-point functions which may give substantial numerical
contributions, especially when $n>N_c$.

The results presented here underscore the importance of rigorous
solutions of the small-$x$ evolution of higher point functions of
Wilson lines. These could be obtained numerically via lattice-gauge
theory techniques along the lines of ref.~\cite{rw,kkrw} where the
small $x$ evolution of the two point function has been studied. For
the two-point function, those authors found only very minor
differences between the JIMWLK and BK (Gaussian $+$ leading $N_c$
approximation) evolution equations. One may expect that due to the
many leading-$N_c$ terms missed by the Gaussian approximation the
differences between the JIMWLK and BK evolution of the $6$-point
function should be substantial (see also
\cite{Dumitru:2010mv,Marquet:2010cf} for other observables where
possible differences between JIMWLK and BK were investigated). Once
(numerical) solutions to these evolution equations become available,
they could also be used to improve present
calculations~\cite{Kharzeev:2004bw,Frankfurt:2007rn,Tuchin:2009nf,Albacete:2010pg}
for dijet production and angular correlations.

\appendix

\section{Forward $q+g$ production in pA collisions} \label{sec:qgXsec}

Here, we reproduce the expression for forward $q+g$ production (with
comparable rapidities) in a valence-quark nucleus collision from
ref.~\cite{Marquet:2007vb}. This expression is to be convoluted with
valence quark distribution functions of a proton (or deuteron) and
with parton $\to$ hadron fragmentation functions in order to obtain a
physical cross section, see refs.~\cite{Albacete:2010pg,Marquet:2007vb}.

The $qA\to qgX$ hard scattering cross section is given by
\bea
\frac{d\sigma^{qA\to qgX}}{d^3k\, d^3q}&=& \alpha_S C_F\,
\delta(p^+\!-\!k^+\!-\!q^+)
\int\frac{d^2x}{(2\pi)^2}\frac{d^2x'}{(2\pi)^2}\frac{d^2b}{(2\pi)^2}
\frac{d^2b'}{(2\pi)^2}\ 
e^{ik_\perp\cdot(x'-x)+i(q_\perp-p_\perp)\cdot(b'-b)} \nonumber\\
& & \sum_{\lambda\alpha\beta}
\phi^{\lambda^*}_{\alpha\beta}(p,k^+,x'-b')\,
\phi^{\lambda}_{\alpha\beta}(p,k^+,x-b)\, \nonumber\\
& & \hspace{-1cm} \left[S_6(b,x,b',x)
-S_4(b,x,b'+z(x'-b')) -S_4(b+z(x-b),x',b')
+S(b+z(x-b),b'+z(x'-b'))\right]~.  \label{qTtoqgX}
\eea
Here, $z\!=\!k^+/p^+$ and $\phi^{\lambda}_{\alpha\beta}(p,k^+,x)$
denotes the amplitude of the $|qg\rangle_0$ Fock state component in
the wave function of a dressed quark to leading order in $\alpha_s$;
the explicit expression is given in ref.~\cite{Marquet:2007vb}. 
Lastly, the various $n$-point functions (target averages) are given by
\be
S_6(b,x,b',x')=
\frac{1}{C_FC_A}\langle\Tr
V_b V^\dagger_{b'} t^d t^c
[U_x U^\dagger_{x'}]^{cd}\rangle\ ,\label{Sqgqg}
\ee
\be
S_4(b,x,b')
=\frac{1}{C_FC_A}
\langle\Tr V^\dagger_{b'} t^c V_b t^d
U^{cd}_x \rangle\ ,\label{Sqgq}
\ee
\be
S(b,b')=\frac{1}{N_c}\langle\Tr
V_b V^\dagger_{b'} \rangle~.\label{Sqq}
\ee

\begin{acknowledgments}
We thank F.~Gelis and Yu.~Kovchegov for useful discussions during the
``High Energy Strong Interactions 2010'' workshop at Yukawa Institute
for Theoretical Physics in Kyoto, Japan.
We gratefully acknowledge support by the DOE Office of Nuclear Physics
through Grant No.\ DE-FG02-09ER41620 and from The City University of
New York through the PSC-CUNY Research Award Program, grants
60060-3940 (A.D.) and 62625-40 (J.J.M.). We further thank the
[Department of Energy's] Institute for Nuclear Theory at the
University of Washington and the Yukawa International Program for
Quark-Hadron Sciences at Yukawa Institute for Theoretical Physics,
Kyoto University, for their hospitality and partial support during the
late stages of this work.
\end{acknowledgments}


\end{document}